\documentclass[11pt]{article}
 \usepackage[dvips]{graphicx}
\begin{document}
\title {A comment on the paper "Deformed Boost Transformations
that saturate at the Planck Scale" by N.B.Bruno,G.Amelino-Camelia,
and J.Kowalski-Glikman}
\bigskip
\author{~ Alex Granik\thanks{Department of Physics, University of the Pacific,
Stockton,CA.95211;~E-mail:agranik@uop.edu}}
\date{}
\maketitle
\begin{abstract}
An alternative ( simplified) derivation of the dispersion relation
and the expressions for the momentum-energy 4-vector $p_i,p_0$
given initially in \cite{NB} is provided. It has turned out that
in a rather "pedestrian" manner one can obtain in one stroke not
only the above relations but also the correct dispersion relation
in $\omega-k_i$ space, consistent with the value of a velocity of
a massless particle. This is achieved by considering the standard
Lorentz algebra for $\omega-k_i$-space. A non-uniqueness of the
choice of the time-derivative in the presence of the finite length
scale is discussed. It is shown that such non-uniqueness does not
affect the dispersion relation in $\omega-k_i$-space. albeit
results in different dispersion relations in $p-p_0$-space
depending on the choice of the definition of the time derivative.
\end{abstract}
\section{Introduction}
In this paper we revisit the  work by N.B.Bruno,G.Amelino-Camelia,
and J.Kowalski-Glikman $^ {\cite{NB}}$ on finite boost
transformations (BAK transformations) following from the
$\kappa$-Poincare algebra. The prime motivation of our return to
these results was that in the space of the boost parameter $\xi$
the commutation relations of the phase portion of the algebra are
pronouncedly non-symmetric.

As a result this leads to somewhat involved exact explicit
expressions for the spatio-temporal components of the
four-momentum as functions of the boost parameter $\xi$. We
$re-derive$ all their results by introducing  Lorentz algebra (a
subalgebra of the original algebra). Here the original components
of the four-momentum \cite{NB} are replaced by the transformed
quantities. The resulting equations are amazingly simple ( in this
sense our work follows the approach of \cite{NB} where the
difficulties encountered in an earlier work \cite{JL} have been
obviated due to a judicious choice of the appropriate
$\kappa$-Poincare algebra) and allow one to find a very simple
form of the explicit expressions obtained in \cite{NB} for the
original quantities as functions of $\xi$, the dispersion equation
for the old and transformed quantities. The latter turned out to
be an exact analog of the respective relation in special
relativity.

The kinematic part of the respective Lorentz algebra requires an
introduction of the transformed boost parameter $z=z(\xi)$
${^{\cite{JG}}}$. This tells us that phase and position sector of
the $\kappa$-Poincare algebra cannot be "diagonalized"
simultaneously.
\section{Finite boost transformations in standard Lorentz algebra and
their relation to the BAK transformations}

Our point of departure is the commutation relations $^{\cite{NB}}$
of $\kappa$-Poincare algebra pertinent to the finite boost
transformations :
\begin{eqnarray}
\label{eq:100}
[N_j,p_r]=i\delta_{jr}(\frac{1-e^{-p_0\lambda}}{2\lambda}+\frac{\lambda}{2}
|\vec{p}|^2)-i\lambda p_jp_r,\nonumber  \\
\lbrack N_j,p_0\rbrack=i p_j
\end{eqnarray}
where $p_0,p_j$ are the temporal and spatial components of the
four-momentum, $\lambda$ is the observer-independent scale,  $j,r
~=~ 1,~2,~3,$ and $N_j$ are the boost generators.

 We introduce new transformed quantities $n_j$, $\omega, k_j$
(where the latter represent the frequency-wave number four-vector)
such that the respective commutation relations are standard
Lorentzian:
\begin{eqnarray}
\label{eq:110}
[n_j,k_r]=i\delta_{jr}\omega \nonumber  \\
\lbrack n_j,\omega\rbrack=i k_j
\end{eqnarray}
In the following we restrict our attention to the boost in the
direction of the axis $1.$ Clearly this does not lead to a loss of
generality since we cam always choose the direction of boost as
the axis $1$. The differential equations corresponding to the
differential representation of $n_1$ are
\begin{eqnarray}
\label{eq:115} \frac{d\omega}{d\xi}=k_1\nonumber \\
\frac{dk_1}{d\omega}=\omega
\end{eqnarray}

For our purposes we rewrite commutation relations (\ref{eq:110})
in a slightly different form:
\begin{eqnarray}
\label{eq:120}
[n_1,|k|]=i\frac{\omega}{|k|}k_1 \nonumber  \\
\lbrack n_1, \omega\rbrack=i k_1
\end{eqnarray}
Therefore the differential representation of the new boost
generator $n_1$ is
\begin{equation}
\label{eq:130} n_1=i\frac{\omega}{|k|}k_1 \frac{\partial}{\partial
|k|}+ik_1\frac{\partial}{\partial \omega}
\end{equation}
The respective characteristic equations to be satisfied by
$(\omega, \vec{k})$
\begin{eqnarray}
\label{eq:140} \frac{d\omega}{d\xi}=k_1\nonumber \\
\frac{d|k|}{d\xi}=\frac{\omega}{|k|}k_1
\end{eqnarray}
 yield the familiar dispersion equation of the special
relativity  in $(\omega, \vec{k})$- space:
\begin{equation}
\label{eq:150}\omega^2-|k|^2=\bar{m}^2=const
\end{equation}
where we call $\bar{m}$ a generalized mass whose relation to the
physical mass will be found later. This relation yields the
following phase and group velocities for a massless case
$\bar{m}$:
$$v_{ph}=v_{g}=1$$
which are consistent with the value of  the velocity of a massless
particle, i.e. $V=1.$

To find the explicit expressions of $p_j$ and $p_0$ as functions
of the boost parameter $\xi$ we write the derivative
$\frac{d\omega}{d\xi}$
\begin{equation}
\label{eq:160}
k_1=\frac{d\omega}{d\xi}=\frac{\partial\omega}{\partial
p_0}\frac{\partial p_0}{\partial
\xi}+\sum\frac{\partial\omega}{\partial p_j}\frac{\partial p_j
}{\partial \xi}
\end{equation}
Using the differential equations satisfied by $(p_0,p_j)$
$^{\cite{NB}}$ we obtain from (\ref{eq:160}) the following
characteristic equations:
\begin{equation}
\label{eq:170} \frac{d\omega}{k_1}=\frac{d p_0}{p_1}=
-\frac{dp_1}{\frac{\lambda}{2}[p_1^2-p_2^2-p_3^2-(1-e^{-2\lambda
p_0})/\lambda^2]}=-\frac{dp_2}{\lambda p_1
p_2}=-\frac{dp_3}{\lambda p_1 p_3}
\end{equation}
From the second and two last equations we obtain the following two
invariants:
\begin{equation}\label{eq:180}
p_2e^{\lambda p_0}=const=k_2,~~~p_3e^{\lambda p_0}=const=k_3
\end{equation}

Inserting (\ref{eq:180}) into the second and third equations of
(\ref{eq:170}), we obtain the following equations in complete
differentials:
\begin{equation}
 dp_0\lambda\frac{p_1^2}{2}e^{\lambda p_0}+e^{\lambda p_0}p_1dp_1+
 d(\frac{p_2^2e^{\lambda p_0}}{2}+\frac{p_3^2e^{\lambda
 p_0}}{2})-\frac{1}{\lambda^2}d[cosh(\lambda p_0)]=0
\end{equation}
yielding another invariant (Casimir operator of the original
algebra) $^{\cite{NB}}$:
\begin{equation}
\label{eq:190} cosh(\lambda
p_0)-\frac{\lambda^2}{2}|p|^2e^{\lambda p_0}=const
\end{equation}\\
To find the $const$ on the right hand side of (\ref{eq:190}) we
notice that if the momentum $|p|=0$ then energy $p_0$ is simply a
particle's mass. This gives the constant's value:
\begin{equation}
\label{eq:195}
 const=cosh(\lambda m)
\end{equation}
Therefore eq.(\ref{eq:190}) becomes:
\begin{equation}
\label{eq:198}
cosh(\lambda p_0)-\frac{\lambda^2}{2}|p|^2e^{\lambda
p_0}=cosh(\lambda m)
\end{equation}
 Equation (\ref{eq:198}) can be recast  into the form given in
\cite{NB}
\begin{equation}
\label{eq:199}
(\frac{2sinh\frac{p_0\lambda}{2}}{\lambda})^2-p^2e^{\lambda p_0}=
(\frac{2sinh\frac{m\lambda}{2}}{\lambda})^2
\end{equation}\\

To find the explicit expressions for $\omega$ and $k_1$ we use
eq.(\ref{eq:150}) to represent $k_1$ in terms of $\omega$ and
eq.(\ref{eq:190}) to represent $p_1$ in terms of $p_0$. Upon
substitution of the results into the first and second equations of
(\ref{eq:170}) we arrive at the following elementary quadratures:
\begin{equation}
\label{eq:200}
\int\frac{d\omega}{\sqrt{\omega^2-D^2}}=\int
\frac{dx} {\sqrt{x^2-2xcosh(\lambda m)+1-\lambda^2(k_2^2+k_3^2)}}
\end{equation}
where $$D^2=\bar{m}^2+k_2^2+k_3^2~~~ and~~x\equiv e^{\lambda
p_0}.$$ Using the fact that $k_2$ and $k_3$ are constant we obtain
from (\ref{eq:200}):
\begin{eqnarray}
\label{eq:210} log( k_1+\omega)=log[p_1e^{\lambda
p_0}+\frac{e^{\lambda p_0}-cosh(\lambda m)}{\lambda}]
\Longrightarrow\nonumber \\
k_1=p_1e^{\lambda p_0},\nonumber \\
\omega=\frac{e^{\lambda p_0}-cosh(\lambda m)}{\lambda}
\end{eqnarray}
This expressions together with eq.(\ref{eq:180}) completely
determine the transformation from $p_j$-$p_0$-space to
$k_j$-$\omega$-space. If we insert (\ref{eq:180}) and
\ref{eq:210}) into the dispersion equation (\ref{eq:150}) and take
into account that for $|k|=0$ $\omega =\bar{m}$ we get the
expression for the reduced mass $\bar{m}$ in terms of the mass $m$

$$\bar{m}=\frac{sinh(\lambda m)}{\lambda}$$

We write all the transformation formulas one more time:
\begin{eqnarray}
\label{eq:215} k_j=p_je^{\lambda p_0},\nonumber \\
\omega=\frac{e^{\lambda p_0}-cosh(\lambda m)}{\lambda},\nonumber\\
\bar{m}=\frac{sinh(\lambda m)}{\lambda}
\end{eqnarray}

We still have to express $\omega$ and $k_1$ in terms of the boost
parameter $\xi$. From eqs.(\ref{eq:115})we obtain:
\begin{eqnarray}
\label{eq:220} \omega=\omega_0cosh\xi+k_{10}sinh\xi\nonumber \\
k_1=k_{10}cosh\xi+\omega_0sinh\xi
\end{eqnarray}
where $k_{10}$ and $\omega_0$ are the values of the respective
quantities at $\xi=0$.

Both these expressions together with the transformation rules
furnished by (\ref{eq:215}) give us (after simple algebra) the
values of $p_0$ and $p_j$ for arbitrary initial ($\xi=0$)
conditions, that is for arbitrary values of $p_{j0}$:
\begin{eqnarray}
\label{eq:230} p_0=\frac{1}{\lambda}log\lbrace\frac{1}{1-\lambda^2
|p|_0^2}[R+\lambda^2|p|_0^2cosh(\lambda m)]cosh\xi+
 \lambda p_{10}[R+ \nonumber \\cosh(\lambda m)]sinh\xi
+cosh{\lambda m}(1-\lambda^2|p|_0^2)]\rbrace\nonumber \\
p_1=\frac{(1-\lambda^2|p|_0^2)R_1sinh\xi+p_{10}[cosh(\lambda
m)+R]cosh\xi} {[R+\lambda^2|p|_0^2cosh(\lambda m)]cosh\xi+\lambda
p_{10}[R+cosh(\lambda m)]sinh\xi+cosh{\lambda
m}(1-\lambda^2|p|_0^2)}\nonumber\\ \nonumber\\
p_{2,3}=\frac{p_{20,30}[cosh(\lambda
m)+R]}{[R+\lambda^2|p|_0^2cosh(\lambda m)]cosh\xi+\lambda
p_{10}[R+cosh(\lambda m)]sinh\xi+cosh{\lambda
m}(1-\lambda^2|p|_0^2)}
\end{eqnarray}
where $$R=\sqrt{sinh^2(\lambda m)+\lambda^2 |p_0|^2}$$
$$R_1=R/\lambda$$ and $$|p|_0^2=\sum p_{j0}^2$$
For a particle initially at rest $|p|_0=0$
and eqs.(\ref{eq:230}) are simplified, yielding the result
discussed in \cite{JG}.

For $\xi\rightarrow\infty$ and $\lambda\xi\neq 0$
eqs.(\ref{eq:230}) become:
\begin{eqnarray}
\label{eq:250} \omega = \frac{\xi}{\lambda} +O(\xi^{-1}),\nonumber \\
p_1=\frac{(1-\lambda^2k^2)R_1+p_{10}[cosh(\lambda
m)+R]}{R+\lambda^2|p_0|^2cosh(\lambda m)+\lambda
p_{10}[cosh(\lambda
m)+R]}+O(e^{-\xi}),\nonumber \\
p_{2,3}\rightarrow 0
\end{eqnarray}\\

We consider now the transformation rules for $x_0$ and $x_i$
introduced in \cite{JG}. Our point of departure would be somewhat
different from \cite{JG}. We begin by directly introducing the
Minkowski distance in $x_0-x_i$-space
\begin{equation}
\label{eq:300} ds^2=dx_0^2-dx_i^2\equiv dx_0^2(1-V_i^2)
\end{equation}
where
\begin{equation}
\label{eq:310} V_i=\frac{dx_i}{dx_0}
\end{equation}

The action according to the conventional prescription is
\begin{equation}
\label{320} S=-a\int ds= -a\int dx_0\sqrt{1-V_i^2}
\end{equation}
The respective lagrangian looks exactly as the standard
special-relativistic lagrangian: $${\cal{L}}=-a\sqrt{1-V_i^2}.$$
The momentum $p_i$ is then
\begin{equation}\label{eq:350}
p_i=\frac{\partial {\cal{L}}}{\partial
V_i}=a\frac{V_i}{\sqrt{1-V_i^2}}
\end{equation}

If we substitute $p_i$ from (\ref{eq:230}), where for  simplicity
sake we take $|p_0|=0,$ we get the following equation:
\begin{equation}
\label{eq:320}
p_i=\frac{\bar{m}sinh\xi}{\lambda\bar{m}cosh\xi+\sqrt{1+\bar{m}^2\lambda^2}}
=a\frac{V_i}{\sqrt{1-V_i^2}}
\end{equation}
Now we make an $ansatz$
\begin{equation}
\label{eq:350} p=\bar{m}\frac{V}{\sqrt{1-V^2}}
\end{equation}
where for simplicity we drop the subscript $i.$  Comparison of
(\ref{eq:320}) and (\ref{eq:350}) yields $a = \bar{m}$, and as a
result the following value of $V$ $^{\cite{JG}}$:
\begin{equation}
\label{eq:400}
V=\frac{sinh\xi}{cosh\xi\sqrt{1+\bar{m}^2\lambda^2}+\sqrt{1+
\bar{m}^2\lambda^2}}
\end{equation}

The most interesting part of this derivation is that the
compatibility of (\ref{eq:300}), (\ref{eq:310}), and
(\ref{eq:400}), requires $$\frac{dx_0}{d\xi}\neq
x;~~~~\frac{dx}{d\xi}\neq x_0$$ The way out of this situation is
to introduce a modified boost parameter $z=z(\xi)$ in
$x-x_0$-space, different from the "bare" boost parameter $\xi$ in
$p-p_0$-space:
$$x=x'coshz+x_0'sinhz;~~~x_0=x_0'coshz+x'sinhz; $$
where prime denote an inertial system $K'$ moving uniformly with
respect to another inertial system $K$.

If initially, a particle was at rest, then $x/x_0=V=tanhz$ and
comparing this expression with eq.(\ref{eq:400}) we find the value
of the boost parameter $z$ $^{\cite{JG}}$:
$$z(\xi)=2tanh^{-1}[e^{-m\lambda}tanh(\frac{\xi}{2})]$$\\

Impossibility to simultaneously introduce standard Minkowski
metric in both $x-x_0$ and $p-p_0$-spaces is due to the fact that
in the respective algebra $x$ and $x_0$ do not commute. Moreover,
the $k-\omega$-space has the Minkowski structure, albeit at the
expense of abandoning the standard relativistic transition
$$\omega=p_0; ~~k=p ~~~ ( in ~ units~ of~\hbar=1). $$ This problem
was discussed in detail in \cite{JM}. In the latter the obtained
dispersion equation differed from the one used in the later
studies (e.g., \cite{NB}, \cite{JG}) and discussed here.\\

However was what not mentioned is an absence of a $unique$
definition of the time derivative in the presence of the minimum
attainable length $\lambda$, even if such a definition is
restricted by a choice of a suitable algebra. This non-uniqueness
is an important factor in the emergence of different dispersion
equations in $p-p_0$-space. Therefore it would look like an
unpleasant fact. However, in  a transition to the correct
dispersion relation $\omega-k$-space this difference (within the
framework of the chosen algebra, $\kappa$-Poincare) does not play
any role, as will be shown below.

In general, one can define such a derivative as follows:
\begin{equation}
\label{500} (\frac{\partial f}{\partial t})_{\lambda}\equiv
\frac{f(t+\alpha\lambda)-f(t-\beta\lambda)}{\gamma\lambda}=
\frac{1}{\gamma\lambda}(e^{\alpha\frac{\partial}{\partial
t}}-e^{-\beta\frac{\partial}{\partial t}})f
\end{equation}
with the following relation between parameters $\alpha, \beta,~
and ~\gamma$:
$$\alpha+\beta=\gamma.$$
For the conventional definition of the derivative the choice of
the parameters is not important, and all different definitions are
equivalent.

In work \cite{JM} the parameters were:
$$\alpha=0,~\beta=i,~\gamma=i$$
In the subsequent studies (e.g., \cite{NB}, \cite{JG}) this choice
was abandoned in favor of
$$\alpha=\beta=\frac{1}{2},~\gamma=1.$$ The respective transition
$\omega\Leftrightarrow p_0, ~k\Leftrightarrow$ is as follows:
$$\alpha=0,~\beta=i,~\gamma=i;~~~\omega^-=\frac{1-e^{-\lambda p_0}}
{\lambda};~~~k=pe^{-\lambda p_0}$$
$$\alpha=\beta=\frac{1}{2},~\gamma=1;~~~\omega^+=\frac{e^{\lambda p_0-1}}
{\lambda};~~~ k=pe^{\lambda p_0}$$

Still the dispersion equations in $\omega-k$-space coincide for
both cases, yielding (\ref{eq:150})$$\omega^2-k^2=\bar{m}^2.$$
Note the ubiquitousness of the factor $e^{\pm p_0\lambda}$,
corresponding to the shifting operator.
\section{Conclusion}
We have provided an alternative (rather elementary) derivation of
the results of the work \cite{NB} using the standard Lorentz group
for the representation of the algebra governing the space
$\omega-k.$ It has turned out that not only the derivation is
simpler, but more important it automatically yields both the
dispersion equation in $p-p_0$-space and $\omega-k$-space, where
the latter is consistent with the velocity of a massless particle.

Interestingly enough, the non-uniqueness of the definition of the
(time) derivative in the environment with the minimum length scale
does not affect the final result, the dispersion equation in
$\omega-k$-space, although the dispersion equations in
$p-p_0$-space are different for different choices of the time
derivative.

It would be interesting to investigate also a situation where one
would take into account not only the explicit inclusion of the
minimum length scale into the commutation relations but also of
the minimum time scale.

\end{document}